\documentclass[aps,pra,twocolumn,amsmath,amssymb,showpacs,superscriptaddress]{revtex4-1}
\usepackage{graphicx}
\usepackage{dcolumn}
\usepackage[mathlines]{lineno}
\usepackage{hyperref}
\usepackage{bm}
\usepackage{epstopdf}
\usepackage{float}
\usepackage{lipsum}
\usepackage{epsfig}
\usepackage{bbold}
\usepackage{color}

\newcommand{\bra}[1]{\ensuremath{\left\langle#1\right|}}
\newcommand{\ket}[1]{\ensuremath{\left|#1\right\rangle}}

\begin{document}

\title{Purity of vector vortex beams through a birefringent amplifier}
\author{Hend Sroor}
\affiliation{School of Physics, University of the Witwatersrand, Private Bag 3, Wits 2050, South Africa}
\author{Nyameko Lisa} 
\affiliation{School of Physics, University of the Witwatersrand, Private Bag 3, Wits 2050, South Africa}
\affiliation{CSIR National Laser Centre, P.O. Box 395, Pretoria 0001, South Africa}
\author{Darryl Naidoo}
\affiliation{School of Physics, University of the Witwatersrand, Private Bag 3, Wits 2050, South Africa}
\affiliation{CSIR National Laser Centre, P.O. Box 395, Pretoria 0001, South Africa}
\author{Igor Litvin}
\affiliation{CSIR National Laser Centre, P.O. Box 395, Pretoria 0001, South Africa}
\author{Andrew Forbes}
\affiliation{School of Physics, University of the Witwatersrand, Private Bag 3, Wits 2050, South Africa}
\email[Corresponding author: ]{andrew.forbes@wits.ac.za}

	
	\date{\today}
	
	\begin{abstract}
\noindent Creating high-quality vector vortex (VV) beams is possible with a myriad of techniques at low power, and while a few studies have produced such beams at high-power, none have considered the impact of amplification on the vector purity.  Here we employ novel tools to study the amplification of VV beams, and in particular the purity of such modes.  We outline a general toolbox for such investigations and demonstrate its use in the general case of VV beams through a birefringent gain medium.  Intriguingly, we show that it is possible to enhance the purity of such beams during amplification, paving the way for high-brightness VV beams, a requirement for their use in high-power applications such as laser materials processing. 
	\end{abstract}
	\maketitle
	

\section{Introduction}
\begin{figure}[H]
	\centering
	\includegraphics[width=1\linewidth,height=12cm]{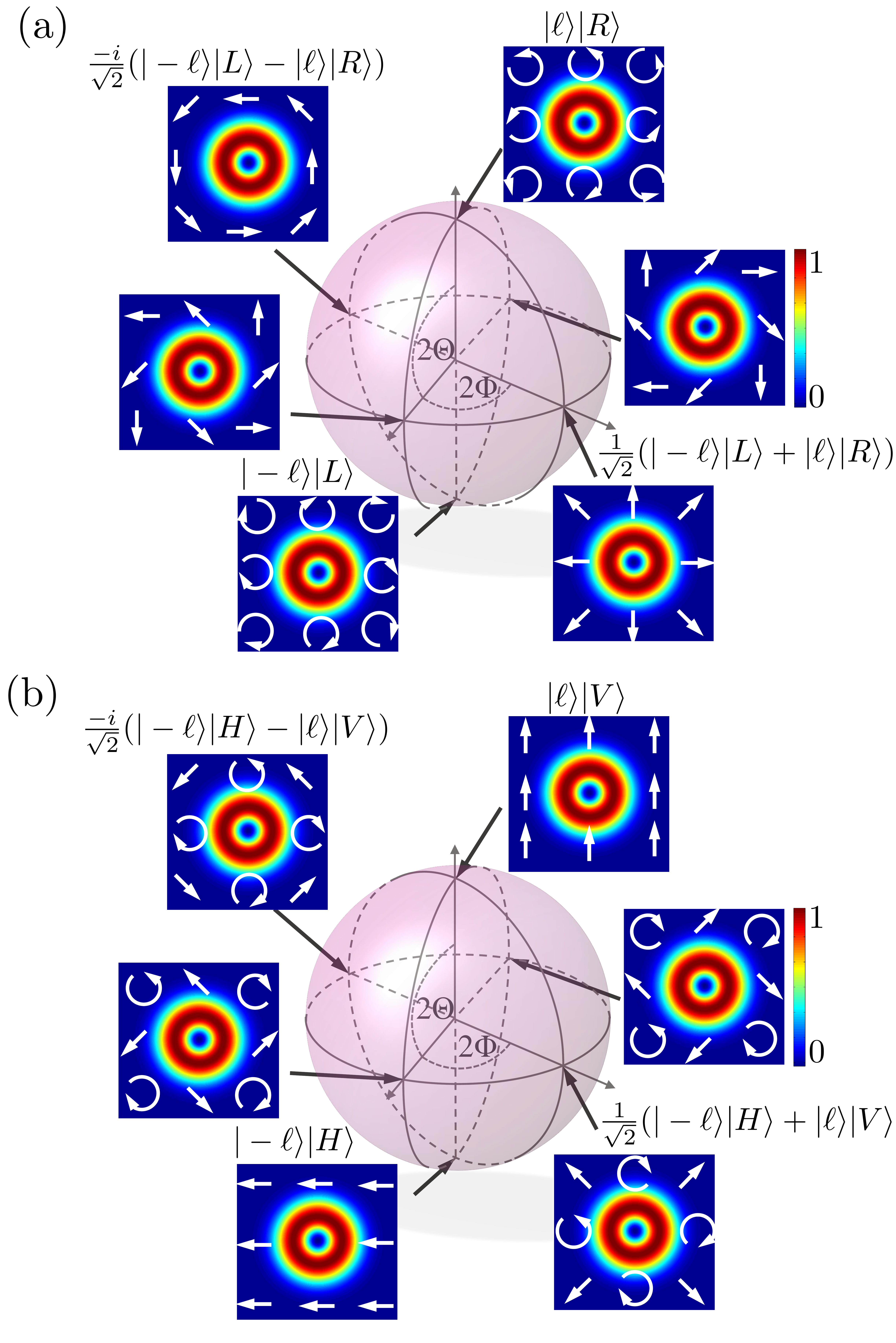}
	\caption{Illustration of (a) a Higher Order Poincar\'e sphere (HOPS) representing VV modes in the left and right polarisation basis and (b) a modified HOPS representing the VV modes in the horizontal and vertical basis. The output beam profile is displayed at different positions on the sphere with arrows indicating polarisation distribution.}
	\label{fig:spheres}
\end{figure}

Laser modes with spatially inhomogenous polarisation states, so called Poincar\'e sphere beams, are highly topical of late \cite{rubinsztein2016roadmap}.  One example of such beams is the well-known cylindrical vector vortex (CVV) beams \cite{Zhan, Brown}. They are characterised by a doughnut-like intensity distribution about a polarisation singularity, and are represented by a point on the Higher Order Poincar\'e Sphere (HOPS) \cite{milione2011higher, milione2012higher} as shown in Fig.~\ref{fig:spheres}(a). Here the poles are scalar circularly polarised orbital angular momentum (OAM) modes with an azimuthally dependent phase of $\exp(\pm i \ell \phi)$, where $\ell$ is the topological charge and $\phi$ is the azimuthal angle, while the CVV beams are found on the equator: radially and azimuthally polarised beams, as well as the two hybrid states, together forming the waveguide modes.  Creating such beams has become common-place in modern laboratories \cite{Niziev, Lu, Naidoo,Grosjean, benuvenu, Stalder, Marrucci, Chen, Han, Andrew, Cardano, Chen, Maurer}, and as a result they have found many applications from imaging, enhanced focussing, metrology as well as fundamental and applied sciences \cite{Novotny, Abouraddy, Donato, Parigi, Li}. 

Almost all studies have employed low-power CVV beams, primarily due to the power handling capability of the custom optical elements used to manipulate them.  Yet applications such as laser-enabled manufacturing have highlighted the need for high power CVV beams \cite{allegre2012laser, golyshev2015effect, weber2011effects}. To address this, CVV beam amplification has become a topic of current research but with only limited studies to date that considered output power of CVV beams from isotropic amplifiers \cite{Loescher, Piehler, Fabien}.  While the output power achievements have been well documented, the purity of the resulting beams remains an open question. As with their scalar counterparts, laser output power must be viewed in the context of the change in beam quality, i.e., the brightness.  It is self-evident that if the vector nature is needed in a particular application, for example, radially polarised light for tight focussing in laser materials processing, then the quality of the radially polarised light will play a significant role in the efficacy of the beam for that application.  Thus we ask the question: how does the vector purity of the VV beam change with amplification?  

\begin{figure*}[ht]
	\centering
	\includegraphics[width=1\linewidth]{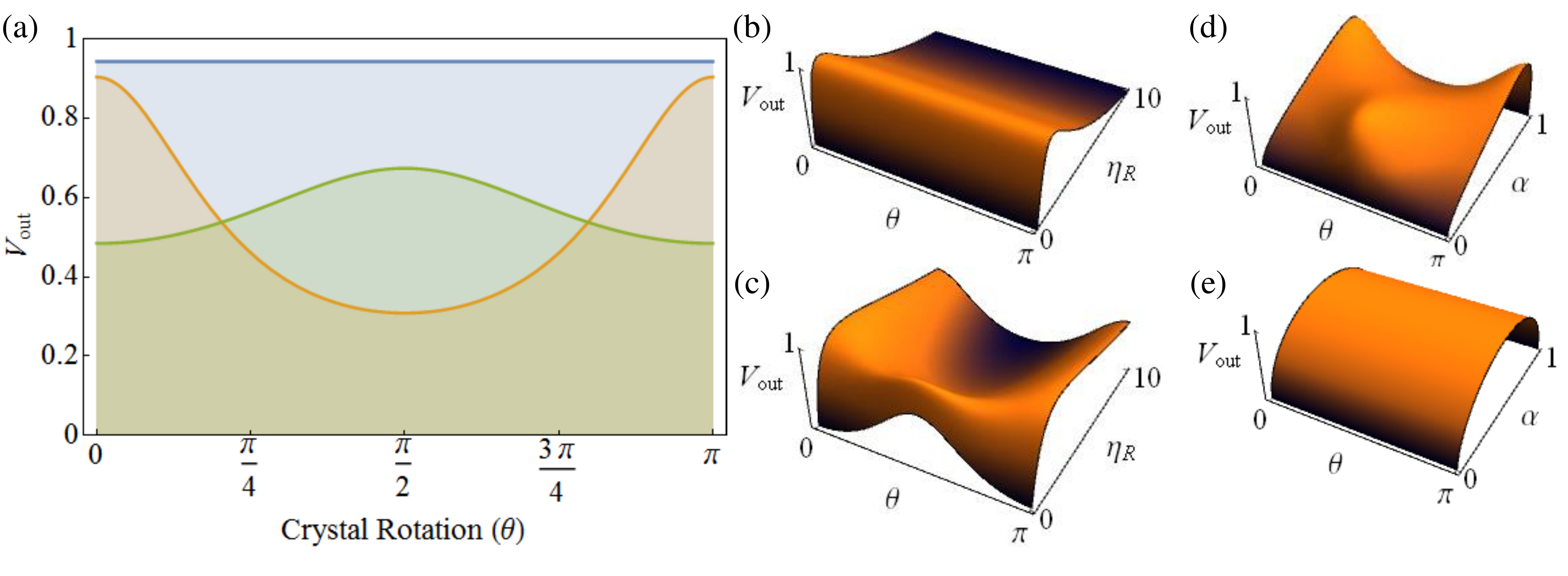}
	\caption{(a) The Vector Quality Factor, $V_{out}$, as a function of crystal angle, $\theta$, for three cases: $\alpha = 0.5$ and $\eta_R = 2$ (blue), $\alpha = 0.8$ and $\eta_R = 0.1$ (orange), and $\alpha = 0.4$ and $\eta_R = 0.1$ (green).  The 3D plots of the parameter space show the richness of the theory, with the input state fixed in (b) and (c) at $\alpha = 0.5$ and $\alpha = 0.7$, respectively.  In (d) and (e) the amplification factor is fixed at $\eta_R = 0.2$ and $\eta_R = 1$, respectively. }
	\label{fig:theory}
\end{figure*}

Here we outline a new approach to studying the amplification of vector beams in general, and CVV beams in particular.  We concentrate on the detection and quantification of the purity of such beams and demonstrate the approach by considering the general case of VV beams through a birefringent gain medium.  We experimentally demonstrate the impact that the amplification process has on such beams, and highlight an intriguing outcome, namely, that it is possible to not only increase the power but also the purity of the amplified mode by judicious choice of gain. This suggests a route to high-brightness vector beams where power and quality are simultaneously increased.
  
\section{Theory}
To begin, let us consider a HOPS beam of the form
\begin{equation}
	U(r,\phi)= \sqrt{\alpha} \exp(i \ell \phi) \space \mathbf{\hat{e}_R} + \exp(i \gamma) \sqrt{1 - \alpha} \space \exp(-i \ell \phi)\space \mathbf{\hat{e}_L},
	\label{eq:eq1}
\end{equation}
\noindent where the vectors $\mathbf{\hat{e}_R}$ and $\mathbf{\hat{e}_L}$ are the right-circular and left-circular polarisation states, respectively, and $\gamma$ is an intra-modal phase.  For example, when $\alpha = 0.5$ and $\gamma = 0$ we have radially polarised light, which has found many applications as indicated earlier.  The vectorness of such modes can be quantified through a quantum measure of entanglement, here measuring the non-separability of the state.  This allows one to define a Vector Quality Factor (VQF), which we denote by

\begin{equation}
	V = 2 |\sqrt{\alpha (1 - \alpha)}|,
		\label{eq:vqf1}
\end{equation}
\noindent with the range of 0 (scalar) through 1 (vector).  Thus the parameter $\alpha$ defines the transverse mode structure, from purely scalar ($\alpha = 1$ or 0) to purely vector ($\alpha = 0.5$), otherwise it is partially vector, while the phase ($\gamma$) only produces a rotation and thus does not affect vector purity.  For this reason we will neglect it in the rest of this treatment.  It is more convenient to express the polarisation vectors in terms of horizontal and vertical states, $\mathbf{\hat{e}_H}$ and $\mathbf{\hat{e}_V}$, respectively, which we shall do from now on in the notation.  Strictly speaking this removes the cylindrical symmetry so that the study is of a more general VV beam represented by a point on the modified HOPS shown in Fig.~\ref{fig:spheres} (b).  Our beam is then

\begin{equation}
	U(r,\phi)= \sqrt{\alpha} \exp(i \ell \phi) \space \mathbf{\hat{e}_H} + \sqrt{1 - \alpha} \space \exp(-i \ell \phi)\space \mathbf{\hat{e}_V}.
	\label{eq:hv}
\end{equation}

When such a beam is passed through an amplifier, the input field transforms to an output field given by (see Appendix)

\begin{equation}
	\tilde{U} =  \left( {\begin{array}{cc}
  a_1 \exp(i\ell\phi) + a_3 \exp(-i\ell\phi)  \\
   a_2 \exp(i\ell\phi) + a_4 \exp(-i\ell\phi)  \\
  \end{array} } \right), \label{Eq:amp}
\end{equation} 

\noindent where we have written the field with the Jones matrix formalism in the horizontal/vertical basis. Without any loss of generality we imagine the case where the amplifier is orientated to have initial amplification factors $\eta_{h0}$ and $\eta_{v0}$ for horizontal and vertical polarisation, respectively (defined at $\theta = 0$).  Now when the crystal is rotated through an angle $\theta$, the state alters and is uniquely determined from the coefficients (see Appendix)

\begin{eqnarray} 
	a_1 &=& \sqrt{\alpha} \left( \sqrt{\eta_{h0}} \cos^2 \theta + \sqrt{\eta_{v0}} \sin^2 \theta \right), \\
	a_3 &=&  \sqrt{1 - \alpha} \cos \theta \sin \theta \left( \sqrt{\eta_{h0}} - \sqrt{\eta_{v0}} \right),
\end{eqnarray}

\noindent and

\begin{eqnarray} 
	a_2 &=&  \sqrt{\alpha} \cos \theta \sin \theta \left( \sqrt{\eta_{h0}} - \sqrt{\eta_{v0}} \right), \\
	a_4 &=& \sqrt{1 - \alpha} \left( \sqrt{\eta_{v0}} \cos^2 \theta + \sqrt{\eta_{h0}} \sin^2 \theta \right), 
\end{eqnarray}

\noindent with a new VQF given by 

\begin{equation}
	V = \frac{2 |a_1 a_4 - a_2 a_3|}{a_1^2 + a_2^2 + a_3^2 + a_4^2} .
		\label{eq:vqf2}
\end{equation}

\begin{figure*}[ht]
	\centering
	\includegraphics[width=\linewidth]{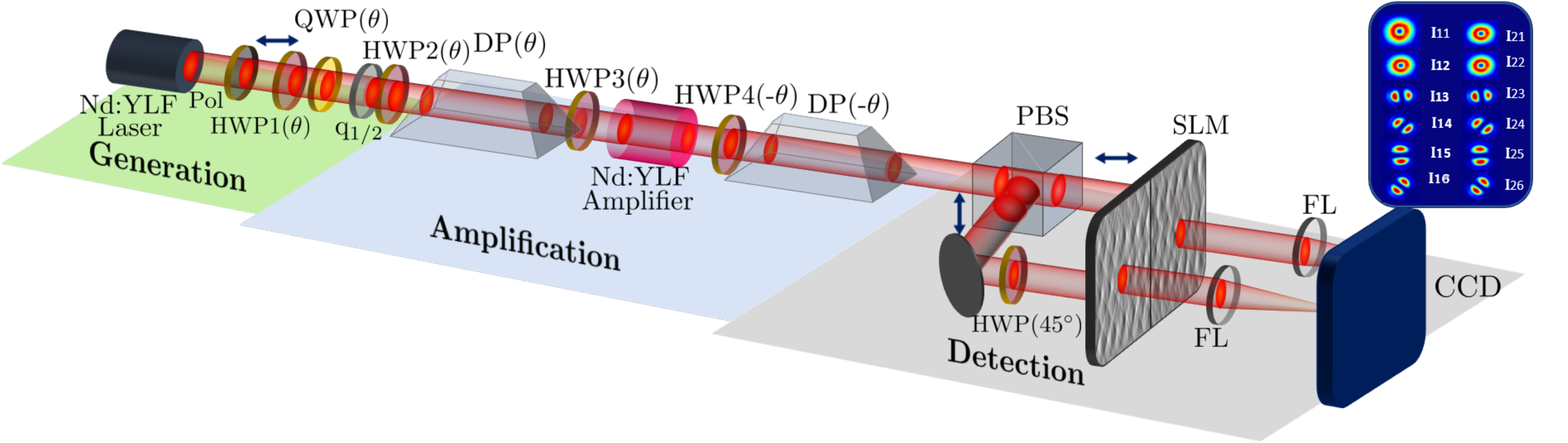}
       \caption{The fundamental Gaussian mode of the source laser was modified in a Generation step with a polariser (Pol) and quarter wave plate (QWP) and passed through a q-plate ($q_{1/2}$) to produce vector beams of variable purity, from 0 to 1.  The resultant beam was passed through the Amplification stage with half wave plates (HWP 3 and 4) dove prisms (DP) either side of the crystal gain medium in order to rotate the beam relative to the crystal axes (effectively rotating the crystal).  Finally, the amplified beam was passed through a Detection system to determine the VQF by a state tomography: after passing through a polarising beam splitter (PBS) each component of the field was measured in the OAM basis by holograms encoded on a spatial light modulator (SLM). The outputs were measured in the Fourier plane using a  charge-coupled device (CCD) camera.}
	\label{fig:intro}
\end{figure*}

For isotropic amplifiers the emission cross-section is independent of polarisation, so $\eta_{h0} = \eta_{v0} = \eta_0$. In this special case the output vector state is identical to the input (purity remains the same) but with an increase in power, i.e., $\tilde{U} = \sqrt{\eta_0} U$. 

 In contrast, birefringent amplifiers generally have different emission cross-sections for each polarisation and thus $\eta_{h0} \neq \eta_{v0}$.   Here, each component of the vector state is amplified differently so that the vector nature (purity) itself changes along with the power content.  Some examples of the effect of a birefringent amplifier are shown in Fig.~\ref{fig:theory}, where the output purity $V_{out}$ is plotted against the the crystal rotation angle $\theta$, the initial amplification ratio $\eta_{R} = \eta_{h0}/ \eta_{v0}$ at $\theta = 0$, and $\alpha$ controls the input purity $V_{in}$.  
 
We see that when the input VV beam is pure ($V_{in} = 1$), the output vector purity is not a function of the crystal rotation angle and is always less than or equal to the input, as seen in the blue line of Fig.~\ref{fig:theory} (a).  This is understandable as a special case where the beam's polarisation structure is rotationally symmetric and thus invariant to the crystal rotation.  When $\alpha \neq 0.5$ this is no longer true, and we note that the vector purity may increase and/or decrease depending on the initial purity and amplification terms.  For example, when the vertical component is initially larger and has higher amplification than the horizontal, shown in the orange curve of Fig.~\ref{fig:theory} (a), the output purity is always less than the input.  Conversely, when the vertical component is initially smaller than the horizontal but has higher amplification, the purity increases as the power weighting of the modes equalise (green curve).  Note that whether or not the purity can be returned to $V_{out} = 1$ depends on the amplification ratio being sufficient to overcome the initial modal mismatch (in power weighting).  This behaviour is further explored in Figs.~\ref{fig:theory} (b) - (e). 

When the initial state is symmetric in polarisation or the amplification ratio is unity, then the purity is independent of the crystal angle, as shown in Figs.~\ref{fig:theory} (b) and (e).  However, when this is not the case the purity can be increased to a maximum of $V_{out} = 1$ or decreased according to the parameters of the initial beam and the medium, illustrated in Figs.~\ref{fig:theory} (c) and (d).  It is this evolution in purity that we wish to uncover experimentally.  

\section{Experimental Set-up}

Our experiment is illustrated in Fig.~\ref{fig:intro}. A horizontally polarised Gaussian beam was produced from a $\lambda =$1053 nm  continuous wave (CW) Nd:YLF laser source with an average power of approximately 164 mW.
\begin{figure}[H]
	\centering
	\includegraphics[width=1\linewidth]{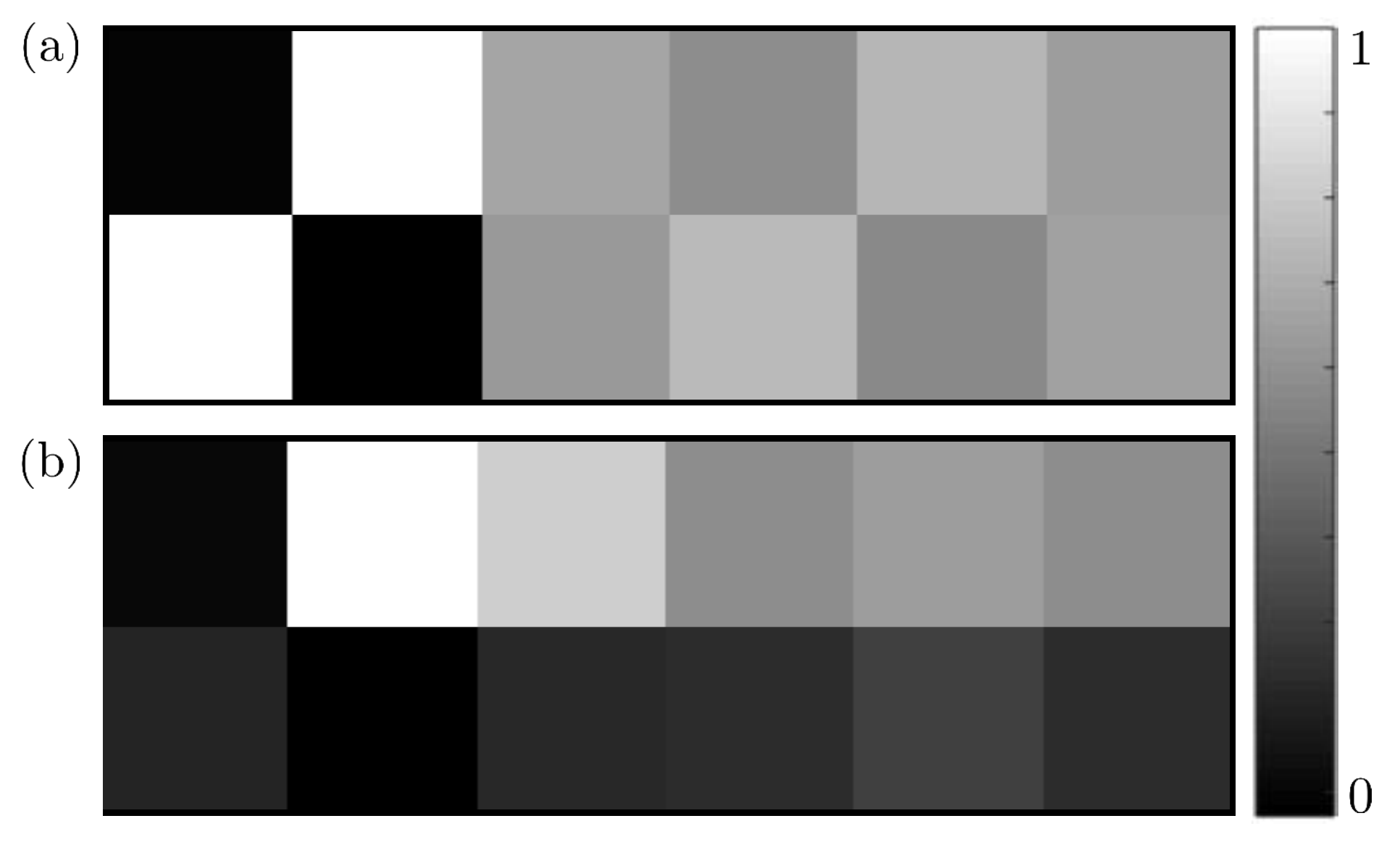}
	\caption{Illustration of the actual intensity measurements $I_{ij}$,
		corresponding to (a) a perfect input VV beam with $V = 1$ and (b) an amplified VV beam with $V = 0.7$.}
	\label{fig:12measure}
\end{figure}
\begin{table}
	\begin{tabular}{c|c| c| c |c| c| c c}
	
		& $\ell=1$ & $\ell=-1$ & $\delta=0$ & $\delta=\pi/2$ & $\delta=\pi$ & $\delta=3\pi/2$& \\ 
		Basis&\includegraphics[width=0.07\linewidth]{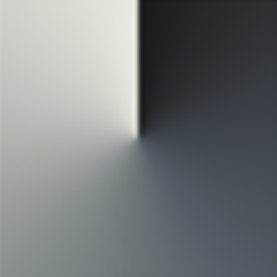}&\includegraphics[width=0.07\linewidth]{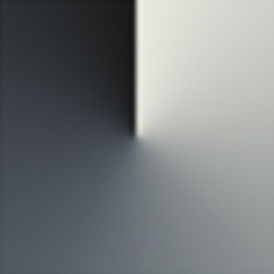}&\includegraphics[width=0.07\linewidth]{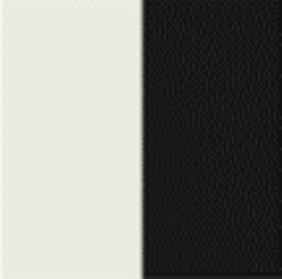}&\includegraphics[width=0.07\linewidth]{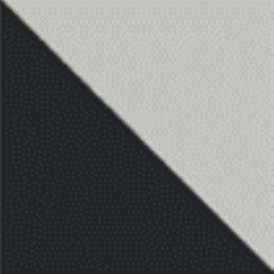}&\includegraphics[width=0.07\linewidth]{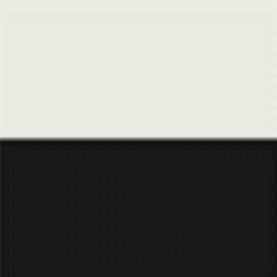}&\includegraphics[width=0.07\linewidth]{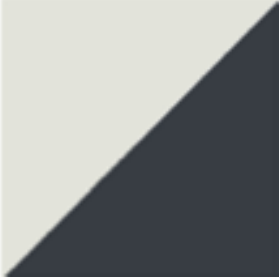}&\includegraphics[width=0.05\linewidth,height=0.6cm]{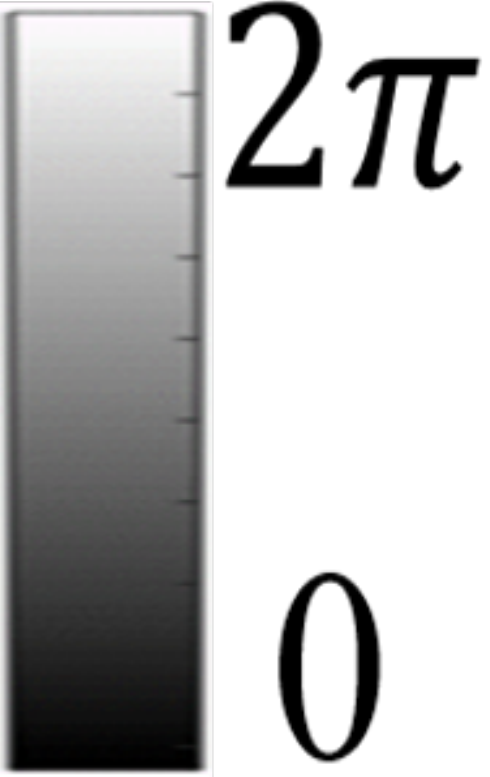}\\ \hline \hline
		$H$& $I_{11}$ & $I_{12}$ & $I_{13}$ & $I_{14}$ & $I_{15}$ & $I_{16}$ &\\ \hline
		$V$& $I_{21}$ & $I_{22}$ & $I_{23}$ & $I_{24}$ & $I_{25}$ & $I_{26}$& \\ \hline
	\end{tabular}
	\caption{The twelve projective measurements to find the unknown intensities, $I_{ij}$, for the calculation of the purity $V$.}
	\label{measurements}
\end{table}
  The beam polarisation was adjusted (with Pol and QWP) and passed through a geometric phase element to create vector beams of variable purity. In this experiment we created VV beams using a q-plate, a birefringent wave plate with an azimuthally varying geometric phase \cite{marrucci2006optical, slussarenko2011tunable, cardano2012polarization}. The q-plate performs spin (polarisation) to orbit (our OAM mode) coupling following the ladder rules: 
$|\ell,L\rangle \rightarrow|\ell+2 q,R\rangle $ and $|\ell,R\rangle \rightarrow|\ell- 2 q,L\rangle $ where $q$ is the topological charge of the q-plate (we have used the Dirac notation for conciseness).  Here, we exploit a q-plate with $q = 1/2$ to produce VV beams with OAM modes of $\ell = \pm 1$.  Our VV beam was passed through an amplifier comprising an uniaxial (birefringent) Nd:YLF crystal (6 mm diameter by 48 mm in length) which was end-pumped by a 805 nm wavelength CW fiber coupled diode laser at 25 W of power. The crystal was mounted within copper blocks, maintained at 293 K by water cooling, with the c-axis initially horizontal, thus defining the $\theta = 0$ position.  Rather than rotate the crystal, which is cumbersome, we performed an identical experiment of rotating the initial beam relative to the crystal with the aid of a dove prism and a half wave plate (HWP3).  Because the subsequent analysis of the beam was polarisation sensitive, a second half wave plate (HWP4) and a dove prism were used to undo this rotation.

\begin{figure}
	\centering
	\includegraphics[width=1\linewidth]{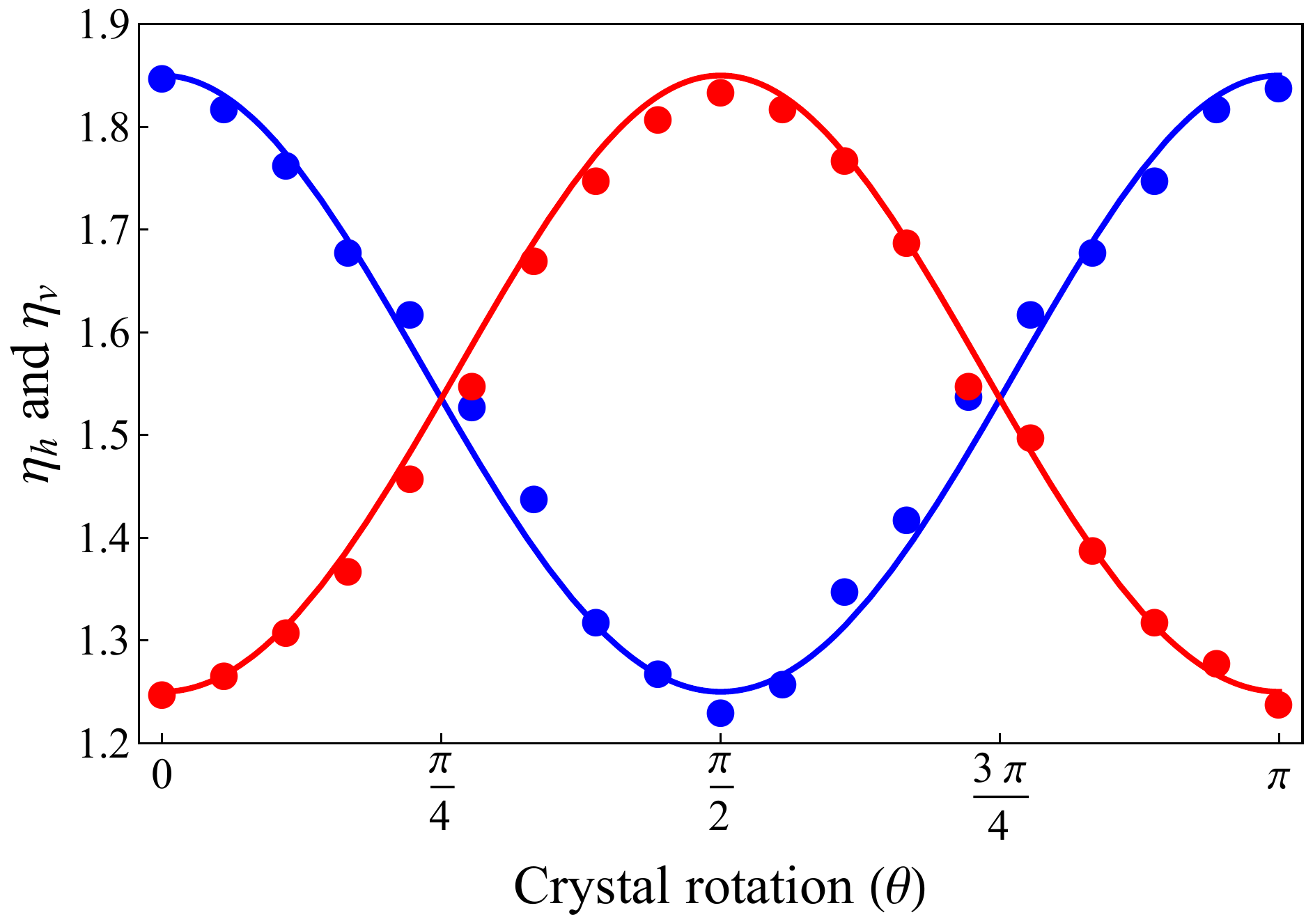}
	\caption{Evolution of the horizontal and vertical amplification factors $\eta_h$ (blue) and $\eta_v$ (red) as a function of the orientation of Nd:YLF crystal, $\theta$.  The points show the experimental data with the theoretical curves overlaid.  The initial amplification factors at $\theta = 0$ are taken as our two parameters, $\eta_{h0}$ and $\eta_{v0}$, for all calculations.}
	\label{fig:amp}
\end{figure}

The VQF was found experimentally by performing a state tomography of the amplified beam \cite{mclaren2015measuring}.  This involved inner product measurements on each polarisation component of the field in order to reconstruct the density matrix, from which the non-separability of the beam, i.e., its vector purity ($V$), could be found from
\begin{equation}
V = \mbox{Re}\left( \sqrt{1- \sum_{i=1}^{3}  S_i^2} \right),
\label{Eq:eq2}
\end{equation}
\noindent where the $S_i$ parameters can be calculated from twelve normalized on-axis intensity measurements \cite{hend} 

\begin{align}
\begin{split}
S_1 = (I_{13}+I_{23})- (I_{15}+I_{25}),
\\
S_2 = (I_{14}+I_{24})- (I_{16}+I_{26}),
\\
S_3 = (I_{11}+I_{21})- (I_{12}+I_{22}).
\end{split}
\end{align}

These twelve measurements $I_{ij}$ were obtained by performing six projection measurements onto two orthognal basis states. For example, if horizontal and vertical polarisation are chosen as basis states, then the six projection measurements are two OAM modes with $\ell =\pm 1$ and four superposition states given by $\exp(i \ell \phi) +\exp(i \delta) \exp(-i \ell \phi)$ with $\delta={0,\pi/4, \pi/2,3\pi/2}$ as in Table.~\ref {measurements}.

\begin{figure}
	\centering
	\includegraphics[width=1\linewidth]{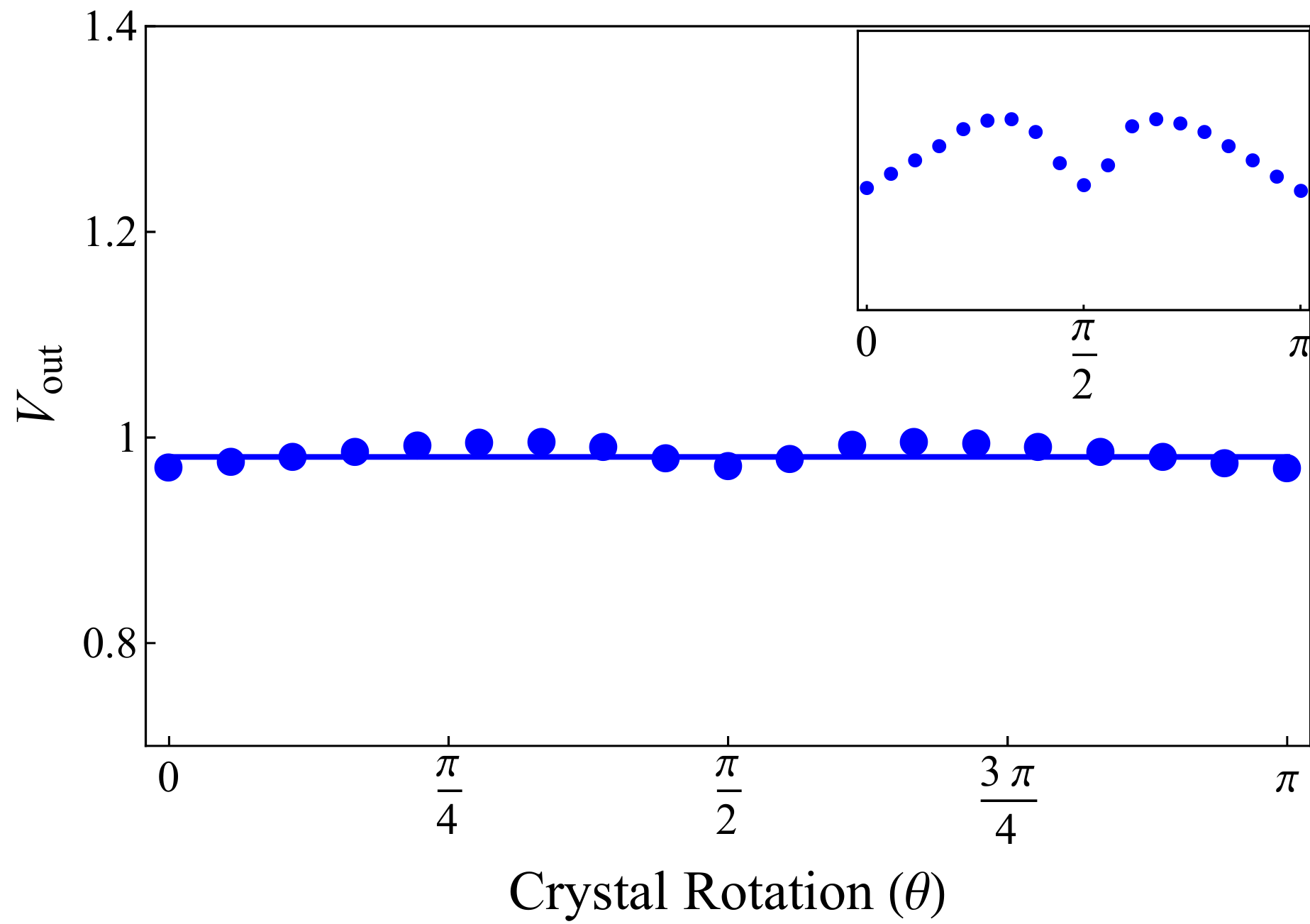}
	\caption{Purity of the output beam over different orientations of the Nd:YLF amplifier for an input purity of $V = 1$.  The theory is shown as the solid curve together with the experimental data points.  The theory predicts $V = 0.99$ and independent of angle, which is corroborated by experiment.  The inset, a zoomed in data sequence, shows a slight modulation due to systematic experimental uncertainty in the form of rotation dependent misalignment, an error in the order of $1\%$. }
	\label{fig:flat}
\end{figure}

This was experimentally realized using a polarising beam splitter (PBS) to spatially seperate the horizontal and vertical polarisation basis states which were directed toward an SLM encoded with the six OAM projections on a single hologram, as shown in Fig.~\ref{fig:intro}. By way of example, the results of the twelve projections for the seed and amplified beams for $\theta =0$ are shown in Fig.~\ref{fig:12measure}(a) and Fig.~\ref{fig:12measure}(b), respectively.

\section{Results}
In order to calibrate the system and ensure correct operation, the purity of the input mode was altered by adjustment of the QWP in Fig.~\ref{fig:intro} and measured with the detection system without any amplification ($\eta_{h0} = \eta_{v0} = 1$).The results confirmed the system performance.  Next, with the gain switched on, the amplification factors were measured as a function of the crystal orientation relative to the input beam.  The experimental results are in good agreement with the theory, exhibiting oscillatory amplification as shown in Fig.~\ref{fig:amp}.  

\begin{figure}
	\centering
	\includegraphics[width=1\linewidth]{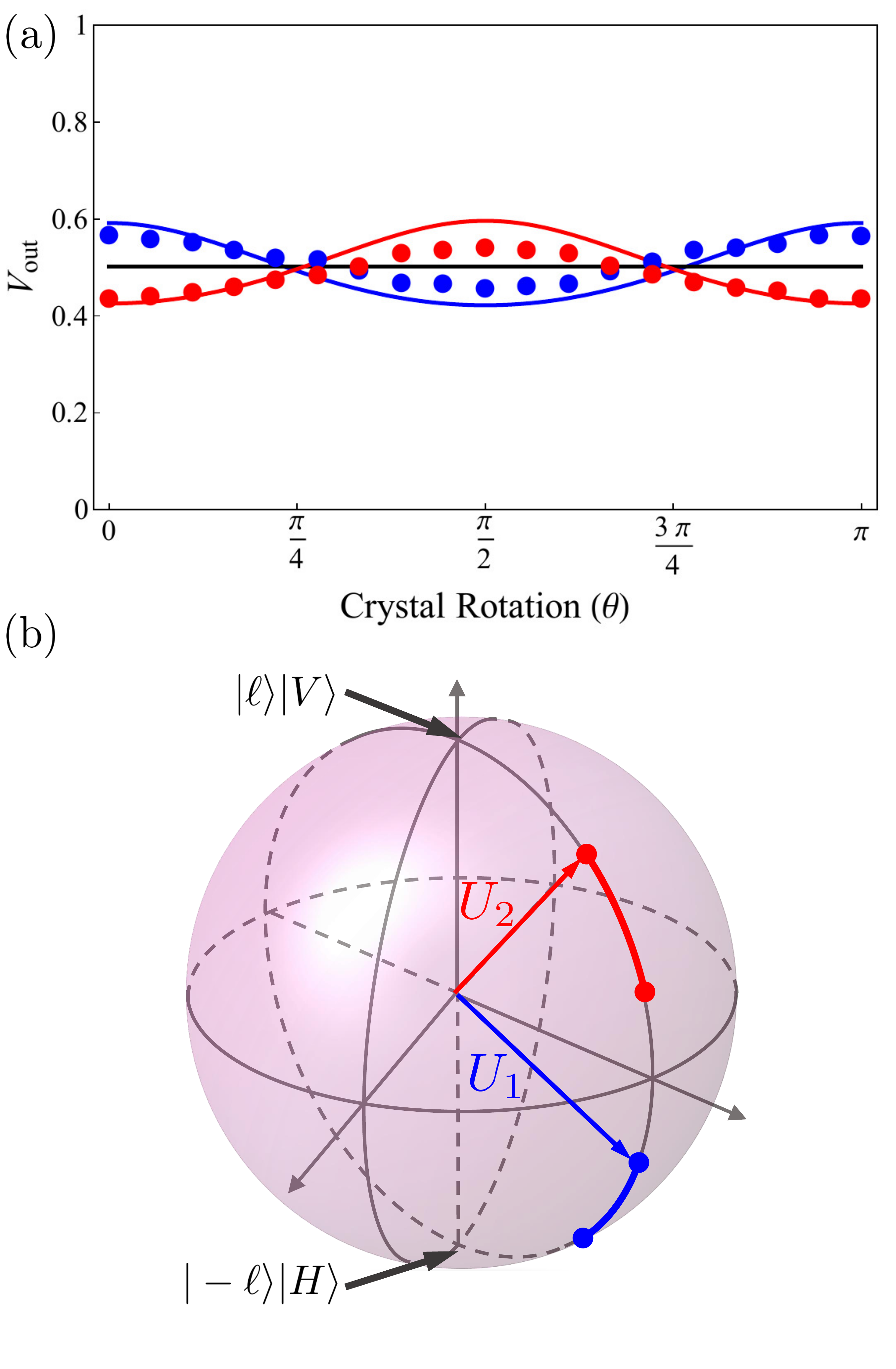}
	\caption{(a) The vector purity of two VV beams, $U_1$ (blue) and $U_2$ (red), as the crystal axis is rotated (for identical $V_{in}=0.5$) .  The purity oscillates about the initial value of $0.5$ (Black solid line) but in some cases is improved due to equalisation of the modal weightings.  (b) The same oscillation in purity is plotted on the HOPS as trajectories for the two beams, show in red and blue.}
	\label{fig:osc}
\end{figure}
This can be understood intuitively: at $\theta = \pi /2$ the input horizontal/vertical components appear swapped, and thus their amplification factors likewise interchange.

Next, a VV beam of $V = 1$ was amplified with the results shown in Fig.~\ref{fig:flat}.  As predicted by theory, the purity of the mode is set by the initial amplification ratio and does not change when the crystal is rotated.  In such a scenario the purity of the initial mode can only \textit{decrease} with amplification unless $\eta_{R} = 1$ (in which case it remains invariant).  
Conversely, the theory suggests that when the initial purity is not perfect, it is possible to increase/decrease the purity depending on the amplification ratio, $\eta_{R}$.  This is confirmed experimentally in Fig.~\ref{fig:osc} (a). 


\begin{figure}
	\centering
	\includegraphics[width=1\linewidth]{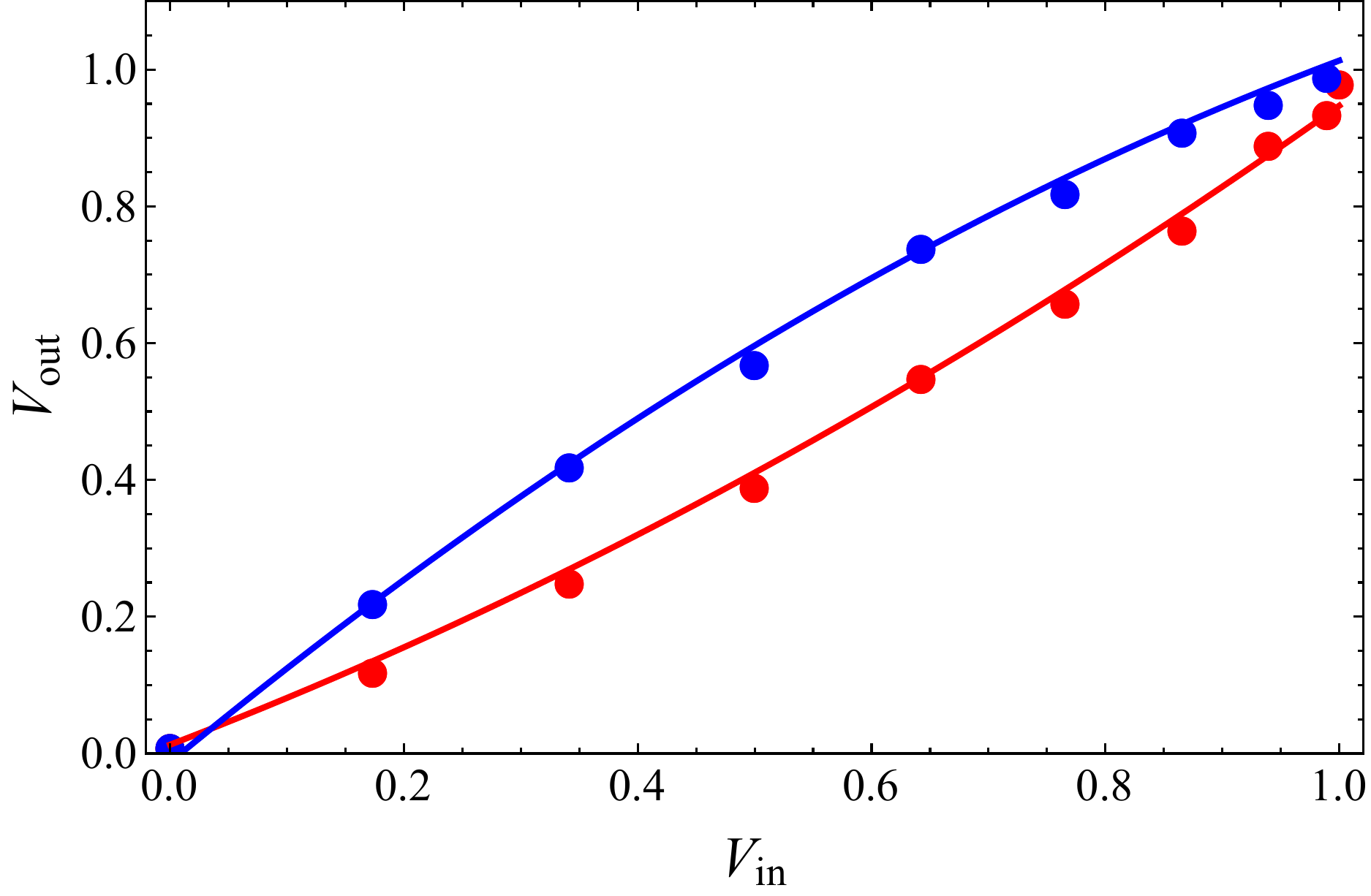}
	\caption{ The change in the purity of the amplified beam ($V_{out}$) with respect to the purity of the injected seed beam ($V_{in}$). By adjusting the QWP, the state of the input field continuously changes. The purity of the amplified beam $V_{out}$ was calculated for QWP angle $\theta \in[0,45]$ (blue dots) and  $\theta \in[0,-45]$ ( red dots).  The experimental data is plotted against the theoretical predictions (solid lines), showing excellent agreement.}
	\label{fig:vqf}
\end{figure}

 Here we selected two initial beams, $U_1$ and $U_2$ from Eq.~\ref{eq:hv}, with $\sqrt{\alpha} = 0.26$ and $\sqrt{\alpha} = 0.965$, respectively, so that both had identical initial purities of $V = 0.5$ before amplification.  We observe the predicted oscillation in purity as a function of the crystal angle, illustrating that the purity could  be improved from the initial value. 

\begin{table}[H]
	\begin{tabular}{c|c| c| c |c| c| c| c| c| c| c}
		QWP($\theta$) & 0$^\circ$  &  5$^\circ$ &  10$^\circ$ & 15$^\circ$ & 20$^\circ$ & 25$^\circ$ & 30$^\circ$ & 35$^\circ$ & 40$^\circ$ & 45$^\circ$ \\  \hline\hline
		$\alpha$&1& 0.99&   0.97 & 0.93 & 0.89 & 0.82 & 0.75 & 0.67 & 0.59 & 0.5 \\ \hline
		$V$& 0.0& 0.17&   0.34 & 0.5 & 0.64 & 0.77 & 0.87 & 0.94 & 0.98 & 1 \\ \hline
	\end{tabular}
	\caption{shows the adjustment of the QWP in order to change the purity of the  intial state}
	\label{measure}
\end{table}

Clearly the mechanism is partial equalisation of the modal weightings in the VV beam.  The amplitude of the oscillation is a function of both the initial purity and of $\eta_R$, the available amplification ratio.  This is illustrated on the modified HOPS in Fig.~\ref{fig:osc} (b), showing how oscillation in purity affects the VV beam position on the sphere. Without any amplification, the beam with $V = 0.5$ is represented by a point on the sphere with $2\Theta = 60^\circ$. After amplification, the position of VV beam $U_2$ (red) oscillates on the modified HOPS from $2\Theta \in [48^\circ,74^\circ]$ while the position for VV beam $U_1$ (blue) oscillates $2\Theta \in [106^\circ,132^\circ]$ as shown in Fig.~\ref{fig:osc}(b).  If the amplification ratio was tuneable (see discussion to follow) then it would be possible to increase the purity from any initial value to unity.  

Finally, the purity $V_{out}$ of the amplified beam was measured with respect to the input beam purity $V_{in}$ while the crystal was set at angle $\theta = 0$. The input beam purity $V_{in}$ was adjusted by continuously rotating the orientation of the QWP in Fig.~\ref{fig:intro} according to the settings summarized in Table ~\ref{measure}.

\section{Amplification of CVV beams}

Next, we study the purity of cylindrical vector vortex (CVV) beams in the form of the four well known waveguide modes: ${|TM\rangle}$,${|TE\rangle}$,${|HE_e\rangle}$ and${|HE_o\rangle}$.  These modes, which include radially and azimuthally polarised light fields, were generated by adjusting HWP1 and HWP2 in Fig.~\ref{fig:intro} according to Table ~\ref{table:CVV}.
\begin{table}[H]
	\begin{tabular}{c|c  c c c}
		& $|TM\rangle$ & $|TE\rangle$ & $|HE_e\rangle$ & $|HE_o\rangle$ \\ 
		Modes&\includegraphics[width=0.19\linewidth]{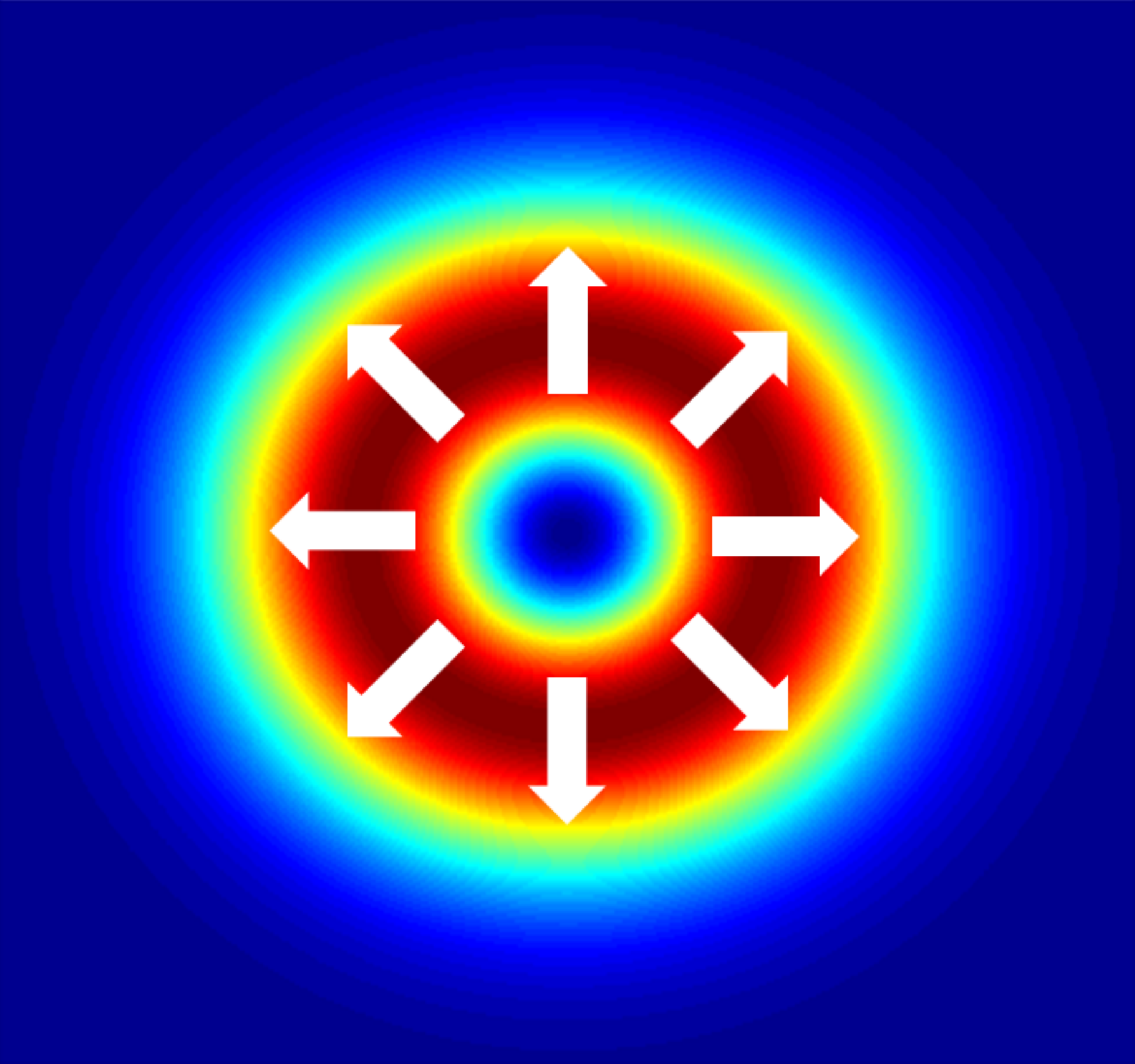}&\includegraphics[width=0.19\linewidth]{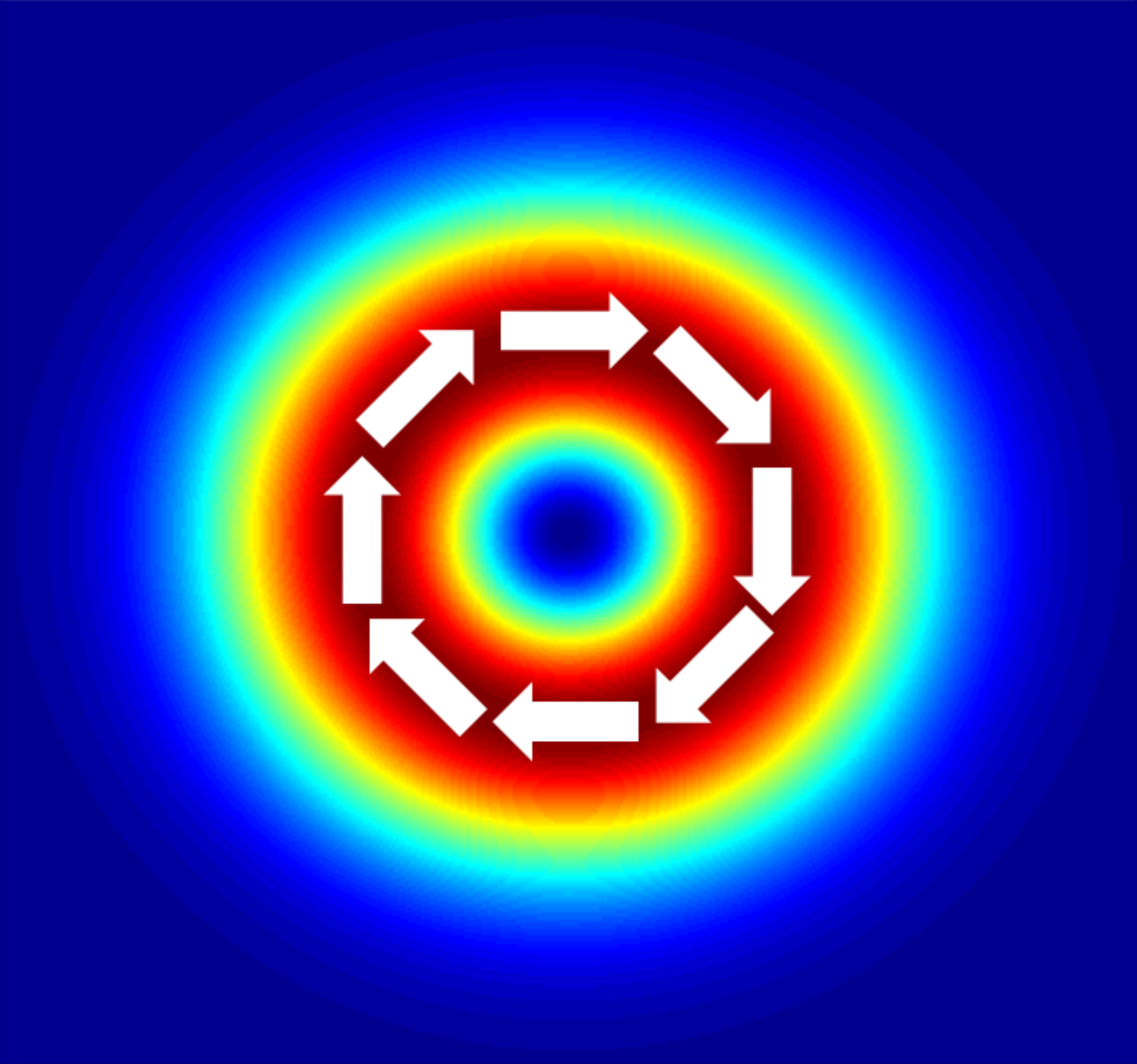}&\includegraphics[width=0.19\linewidth]{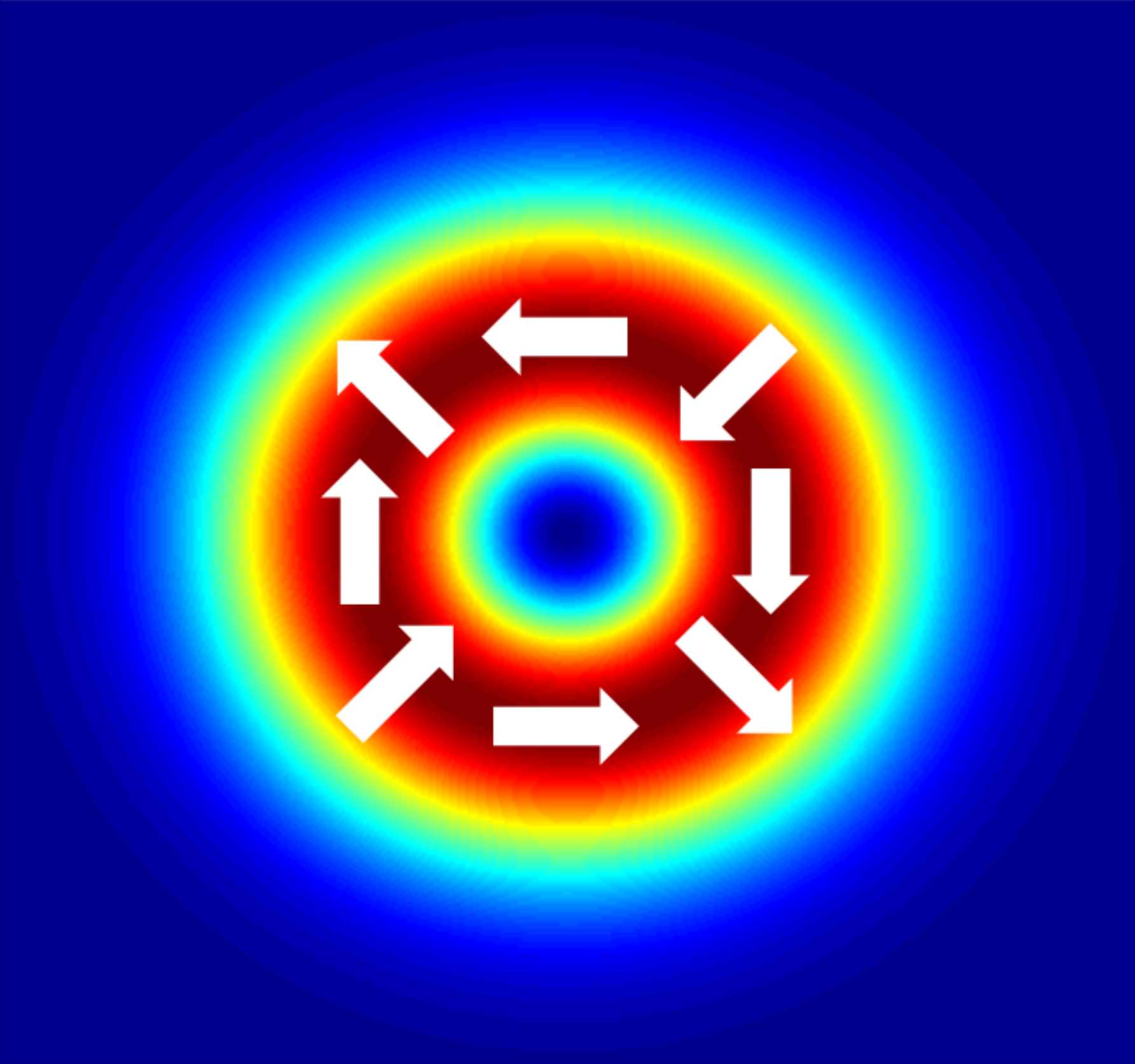}&\includegraphics[width=0.19\linewidth]{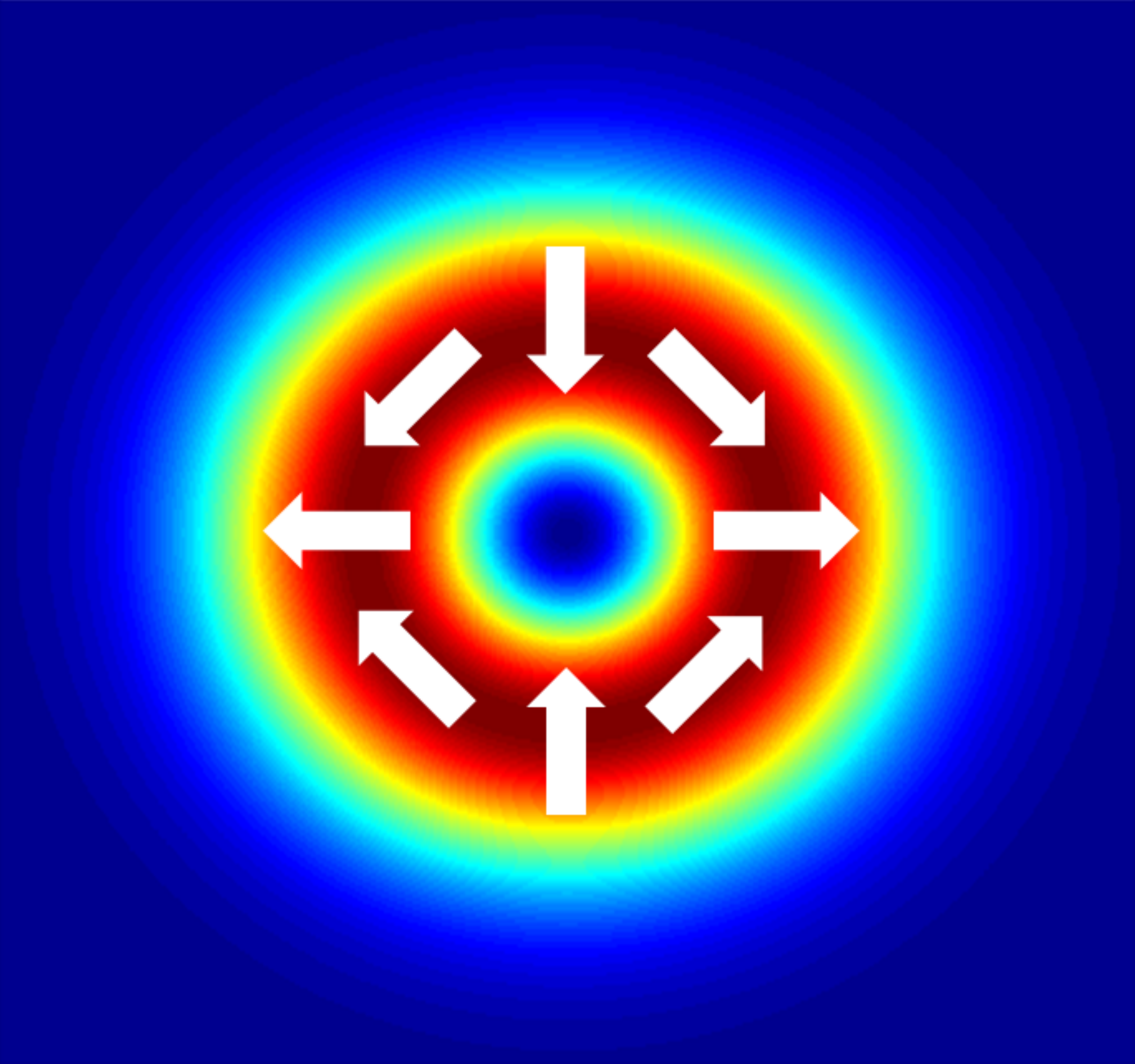}\\ \hline \hline
		HWP1($\theta$)& $0^\circ$ &$45^\circ$ & $0^\circ$ & $45^\circ$  \\ \hline
		QWP($\theta$)& $45^\circ$ &$45^\circ$ & $45^\circ$ & $45^\circ$  \\ \hline
		HWP2($\theta$)&$0^\circ$ &$0^\circ$ & $45^\circ$ & $45^\circ$  \\ \hline
	\end{tabular}
	\caption{Required settings for HWP1, QWP and HWP2 in order to generate pure ${|TM\rangle}$, ${|TE\rangle}$, ${|HE_e\rangle}$ and ${|HE_o\rangle}$ modes. The output CVV beam in each case would have a purity $V = 1$ as these are maximally non-separable, or ideal vector vortex beams.}
	\label{table:CVV}
\end{table}

\begin{figure}
	\centering
	\includegraphics[width=1\linewidth]{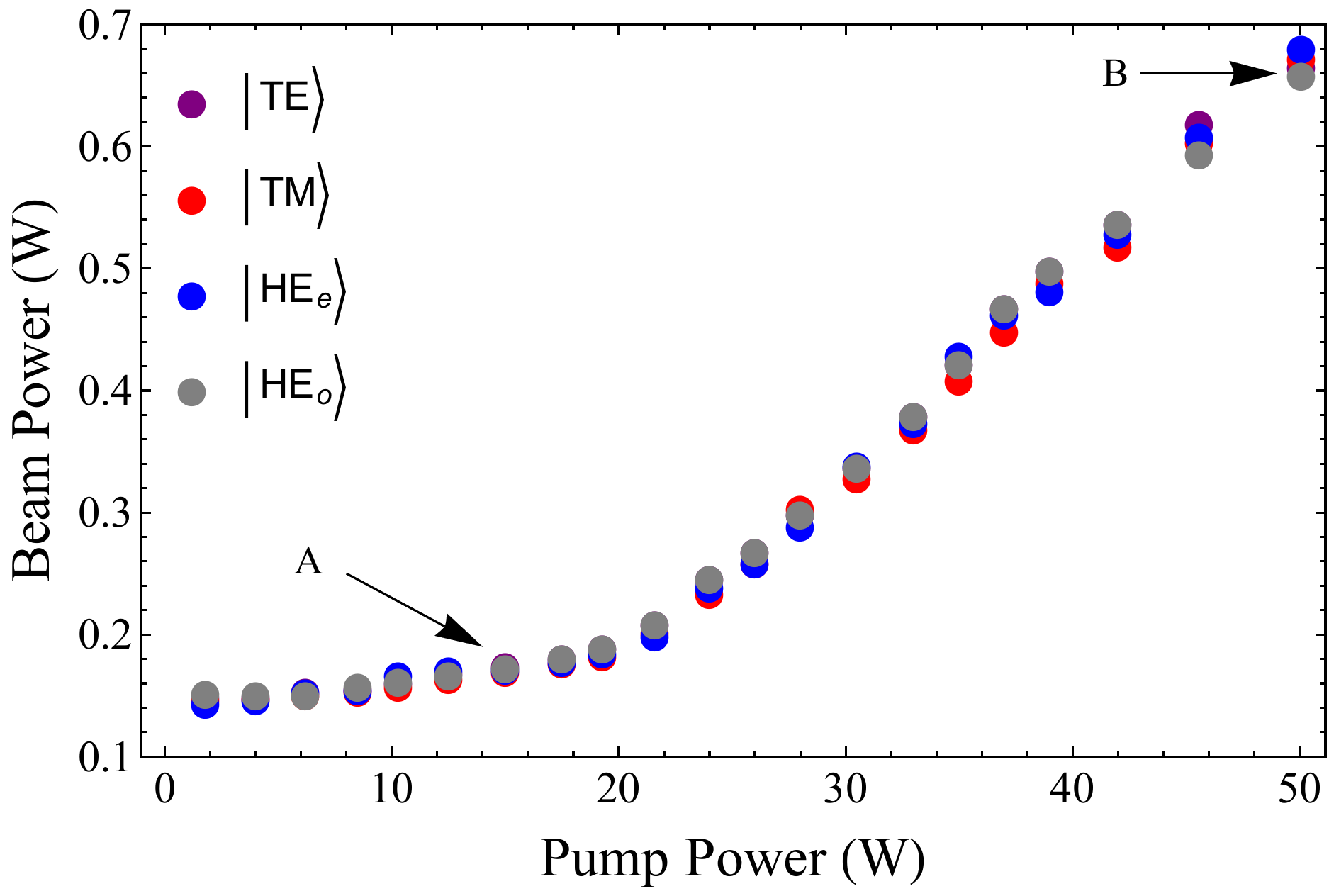}
	\caption{The output power for the CVV beams as a function of the pump power for a given injected power of 164 mW. Point (A) represents the threshold point for the amplifier.}
	\label{fig:power}
\end{figure}

The seed CVV beams were amplified through the Amplification stage as in Fig.~\ref{fig:intro}, with the output of each following a trend in purity as of that in Fig.~\ref{fig:flat}.  To see how the purity changed with amplification, we measured the output power for each CVV beam as a function of the amplifier pump power, with the results shown in Fig.~\ref{fig:power}.   At the threshold power of the amplifier ( point A ), the purity of the amplified CVV beam did not change (equal to the seed beam, $V = 1$).  Well after the threshold point (point B), the purity was measured to be $V = 0.988$ for ${|TM\rangle}$,  $V = 0.982$ for ${|TE\rangle}$, $V = 0.986$ for ${|HE_e\rangle}$ and  $V = 0.983$ for$ {|HE_o\rangle}$.  In other words, the purity of the modes was virtually unchanged.

\section{Discussion and Conclusion}

The approach we have outlined here is general, with OAM modes used by way of example only. This example has been chosen partly due to the ease of creation of such modes and partly because of their topical nature in as far as applications are concerned.  A change of notation from OAM modes to general modes, i.e. $\exp(i \ell \phi) \rightarrow M_{h}$ and $\exp(-i \ell \phi) \rightarrow M_{v}$ will only require a change to the measurement holograms for the tomography without any change to the approach and theory, which is general.  We have explicitly selected the horizontal and vertical basis so that symmetry does not cloud the general behaviour.  What we predict and observe is that the purity of vector modes is an important aspect to consider when amplifying such beams, and that in fact the purity may be improved with careful choice of parameters.  For example, it will be possible by a sequence of crystals, judiciously rotated relative to one another, to overcome any degradation in purity due to the desired amplification.  Although the measured output powers were in the range of $<$ 1 W, we stress that this is a study on purity and not on optimising the output power.  We believe that this work provides an important platform for realising high brightness vector beams with high power and high purity.

In conclusion, we have introduced a novel approach with which to study the amplification of vector beams.  We have shown the impact of birefringent gain on vector purity with non-birefringent gain a special case of our results.  We provide the theory and measurement toolbox for practical use and confirm its validity through experiment under a range of conditions.  Importantly, our results highlight a roadmap towards high brightness vector beams for industrial applications.

\section*{Acknowledgements}
We would like to thank Lorenzo Marrucci for providing the q-plates.


%

\section*{Appendix}

\noindent Here we outline the approach to modelling the observed behaviour.  Let $U$ be expressed as a Jones vector in the horizontal/vertical basis, so that

\begin{equation}
	U =  \left( {\begin{array}{cc}
   \sqrt{\alpha} \exp(i\ell\phi)  \\
   \sqrt{1 - \alpha} \exp(-i\ell\phi)  \\
  \end{array} } \right), \nonumber
\end{equation} 

\noindent which in the Dirac notation reads
\begin{equation}
	\ket{U} = \sqrt{\alpha} \ket{\ell,H} + \sqrt{1 - \alpha} \ket{-\ell,V}. \nonumber
\end{equation} 
\noindent Now we assume that the action of the gain is to modify the modal weightings so that the transfer matrix, $M$, may be written as
\begin{equation}
	M =  \left( {\begin{array}{cc}
   \sqrt{\eta_{h0}}  & 0  \\
   0 & \sqrt{\eta_{v0}}  \\
  \end{array} } \right), \nonumber
\end{equation} 
\noindent where $\eta_{h0}$ and $\eta_{v0}$ represent intensity enhancements when the crystal is orientated at an initial angle $\theta = 0$.  Now we rotate the crystal, transforming the input state to a new output state given by
\begin{equation}
	\tilde{U} = R(\theta) \cdot M \cdot R(-\theta) \cdot U, \nonumber
\end{equation} 
\noindent where $R(\theta)$ is the well-known rotation matrix.  Thus  

\begin{equation}
	\tilde{U} =  \left( {\begin{array}{cc}
   \cos \theta  & -\sin \theta  \\
   \sin \theta & \cos \theta \\
  \end{array} } \right) 
  \left( {\begin{array}{cc}
   \sqrt{\eta_{h0}}  & 0  \\
   0 & \sqrt{\eta_{v0}}  \\
  \end{array} } \right) 
  \left( {\begin{array}{cc}
    \cos \theta  & \sin \theta  \\
  - \sin \theta & \cos \theta \\
  \end{array} } \right) \cdot
 U, \nonumber
\end{equation} 

\noindent or
\begin{equation}
	\tilde{U} =  \left( {\begin{array}{cc}
  a_1 \exp(i\ell\phi) + a_3 \exp(-i\ell\phi)  \\
   a_2 \exp(i\ell\phi) + a_4 \exp(-i\ell\phi)   \\
  \end{array} } \right), \nonumber
\end{equation} 
\noindent which in Dirac notation reads
\begin{equation}
	\tilde{\ket{U}} = a_1 \ket{\ell,H} + a_2 \ket{\ell,V} + a_3 \ket{-\ell,H} + a_4 \ket{-\ell,V}. \nonumber
\end{equation} 
\\
This is a pure state but not always separable.  To calculate the purity we first normalise the coefficients so that $\sum_{i}|a_i|^2 = 1$ and then calculate the purity from the density matrix of the state, $\rho$, following $V = \sqrt{1 - \mbox{Tr}(\rho)}$ with $\rho = \tilde{\ket{U}}\tilde{\bra{U}}$ and where Tr is the trace operator.  This results in an analytical expression given by
\[
V = 2|a_1 a_4 - a_2 a_3|,
\]
\noindent with normalised coefficients or 
\[
V = \frac{2|a_1 a_4 - a_2 a_3|}{\sum_{i}|a_i|^2},
\]
\noindent for unnormalised coefficients, as it would be in the experiment.

\end{document}